\documentstyle[12pt,twoside,fleqn,espcrc1,epsfig]{article}

\newcommand{\AmS}{{\protect\the\textfont2
  A\kern-.1667em\lower.5ex\hbox{M}\kern-.125emS}}

\title{Nuclear absorption and anomalous $J/\psi$ suppression in Pb+Pb
collisions}

\author{A. K. Chaudhuri
\address
{Variable Energy Cyclotron Centre\\1-AF, Bidhan Nagar, Kolkata-700 064}}

\begin{document}

\maketitle

\begin{abstract}
We discuss $J/\psi$ suppression in a QCD based nuclear absorption
model.  Centrality  dependence of $J/\psi$ suppression in S+U and in
Pb+Pb collisions are explained in the model. However,  the  model
fails  to  explain  the  centrality  dependence  of  $\psi\prime$
suppression.   $\psi\prime$   suppression  in  S+U  or  in  Pb+Pb
collisions require additional suppression. Additional  suppression
of  $\psi\prime$, due to hadronic comovers or due to QGP formation
could not be distinguished in Pb+Pb collisions. We then show  that
the   centrality  dependence  of  the  ratio,  $\psi\prime$  over
$J/\psi$, could possibly distinguish two scenario  (e.g.  QGP  or
hadronic comover) at RHIC energy. \end{abstract}

\section{INTRODUCTION}

$J/\psi$ suppression is recognized as one of the promising signal
of  the deconfinement phase transition. Due to screening of color
force, binding of $c\bar{c}$ pair into a $J/\psi$ meson  will  be
hindered,  leading to the so called $J/\psi$ suppression in heavy
ion  collisions  \cite{ma86}.  NA50  collaboration   \cite{na50a}
observed anomalous suppression (suppression beyond 'conventional'
nuclear  absorption)  in  158  GeV/c  Pb+Pb  collisions. 
Several 
authors have explained the 'anomalous $J/\psi$ suppression' with or
without the assumption of QGP formation \cite{bl00,ca00,ch01,qiu98,ch02}.
In \cite{ch02} it was shown that a QCD based nuclear absorption model
could explain the anomalous $J/\psi$ suppression in Pb+Pb collisions.
The model also explain the centrality dependence  of
$p_T$  broadening  of  $J/\psi$  in Pb+Pb collisions \cite{ch02a}.
Recently, in QM2002, NA50 collaboration presented the preliminary
analysis of 2000 Pb+Pb run \cite{na50b}. They also presented  the
analysis  of high statistics pA data \cite{na50c}, implying 
$J/\psi$-nucleon absorption cross  section, $\sigma^{J/\psi N}_{abs}$ 
=$4.4\pm  1$mb,  much
less  than  the  earlier estimate of 6-7 mb. The latest NA50 data
\cite{na50b,na50c}  are  also  well  explained  in  the QCD based
nuclear absorption model \cite{ch03}. 

Here, after a brief description of the QCD based nuclear
absorption model, we have presented our recent analysis  
of 
NA38/NA50 data  on the centrality dependence of $J/\psi$ and $\psi\prime$ 
suppression in S+U   and in Pb+Pb  collisions. Details of the
analysis could be found in \cite{ch03a}.

\section{$J/\psi$ SUPPRESSION IN QCD BASED NUCLEAR ABSORPTION
MODEL}

Details   of  QCD  based  nuclear  absorption  can  be  found  in
\cite{qiu98,ch02}. $J/\psi$ production is assumed to be two  step
process,   (i)  production  of  $c\bar{c}$  pair,  perturbatively
calculable and (ii) formation of  $J/\psi$  meson,  intrinsically
non-perturbative   process,  which  is  parameterized.  In  pA/AA
collisions, the produced pair interact with nuclear medium, which
increases the relative 4-square momentum of the pair. Some of the
pairs can gain enough momentum to cross the  threshold  for  open
charm  production.  Charmonium  production  is then reduced in AA
collisions. Relevant parameter, gain in the 4-square momentum per
unit length ($\varepsilon^2$=0.185 $GeV^2/fm)$ was obtained  from
a  fit to the high statistics pA data on the total $J/\psi$ cross
section  \cite{na50c}.  The  other  parameter   of   the   model,
$B_{\mu\mu}\frac{\sigma^{J/\psi}_{NN}}{\sigma^{DY}_{NN}}$,    was
obtained from  a  fit  to  NA38  S+U  data \cite{su_jpsi}. The same
value was  used  for the latest NA50 Pb+Pb data \cite{na50b}.
  In Fig.1a and 1b, our
model calculations are compared with NA38/NA50 data  on  $J/\psi$
suppression.  Data  are well explained. It can be seen that there
is no scope for additional  suppression  either  due  to  comover
interaction or due to QGP formation.

\begin{figure}[h]
\centerline{\psfig{figure=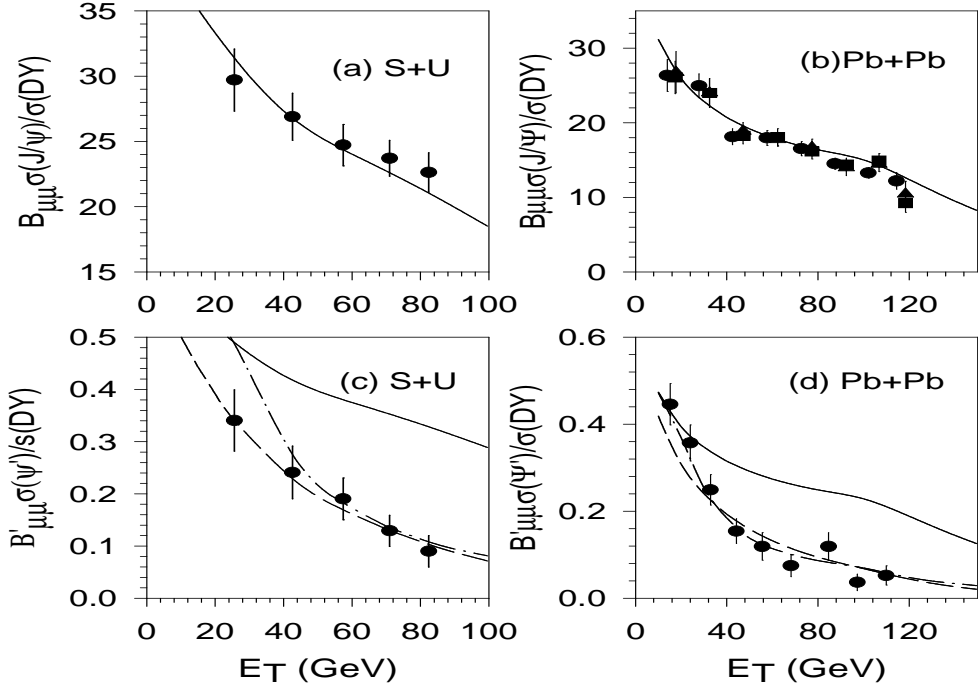,height=12.5cm,width=13cm}}
\vspace{-3.5cm}  \caption{Centrality  dependence of the ratio,
 $J/\psi$ over
Drell-Yan  and $\psi\prime$ over Drell-Yan, in S+U and Pb+Pb collisions. The solid lines are the QCD based nuclear absorption
model calculation.   The dashed
and dash-dotted  lines, for $\psi\prime$ suppression,  are  obtained  with  nuclear+comover  and
nuclear+QGP suppression respectively. } \end{figure}

\vspace{-1cm}
\section{$\psi\prime$ SUPPRESSION}

   It  is
experimentally observed that in pA  collisions,  A-dependence  of
$\psi\prime$     suppression     is     the     same    as    for
$J/\psi$\cite{na50c,e777}. The  phenomena  is  explained  in  the
color  octet  model  \cite{octet}, where $c\bar{c}$ attach with a
collinear gluon  to  neutralise  its  colour.  The  pre-resonance
($c\bar{c}g$)  state  then  transforms  in  to  a  $J/\psi$  or a
$\psi\prime$. In pA collisions,  nuclear  medium  sees  only  the
pre-resonance  state leading to similar A-dependence for $J/\psi$
and $\psi\prime$. In the  QCD  based  nuclear  absorption  model,
nuclear  suppression  of  $J/\psi$ and $\psi\prime$ is due to the
same mechanism, i.e. gain in the relative  4-square  momentum  of
the $c\bar{c}$ pair. The parameter relevant for the A-dependence, the
4-square momentum gain per unit length $\varepsilon^2$, should be
same,  regardless  of  the final state, $J/\psi$ or $\psi\prime$.
Naturally,   $J/\psi$   and   $\psi\prime$  should  show  similar
A-dependence.

In  Fig.1c  and  1d,  NA50/NA38  data \cite{na50a,su_jpsi} on the
centrality dependence of the ratio,  $\psi\prime$  over  DY,  are
shown.  The  solid  lines  are the fit, obtained in the QCD based
nuclear  absorption  model.  Clearly,  the  model  predict   less
absorption   than  the  data  exhibit.  Thus  in  AA  collisions,
additional suppression mechanism is operative for $\psi\prime$'s,
which is absent in pA collisions. The additional  suppression  of
$\psi\prime$   could   be   due   QGP   formation   following  a
deconfinement phase  transition  or  due  to  hadronic  comovers
(unlike  in  pA  collisions,  large number of
secondaries are produced in  AA  collisions).

To  take  into  account  the  suppression  due  to QGP formation,
following Blaizot {\em  et  al.}  \cite{bl00},  we  introduce  an
additional  suppression  factor  ($S_{QGP}$)  such  that  above a
threshold density, $n_c$, all  the  $\psi\prime$  are  dissolved.
Threshold   density  ($n_c$)  is  obtained  from  a  fit  to  the
$\psi\prime$ suppression data \cite{ch03a}. Similarly, to account
for  the  comover  suppression,   suppression   factor   due   to
comover-$\psi\prime$  interaction  ($S_{co}$)  was introduced. We
calculate $S_{co}$  following  \cite{ga96}.  $S_{co}$  involve  a
parameter,    $\psi\prime$-comover   absorption   cross   section
($\sigma_{co}$), which again is obtained from a fit to the experimental
$\psi\prime$ data.

\begin{figure}[htb]
\begin{minipage}[t]{70mm}
\epsfig{figure=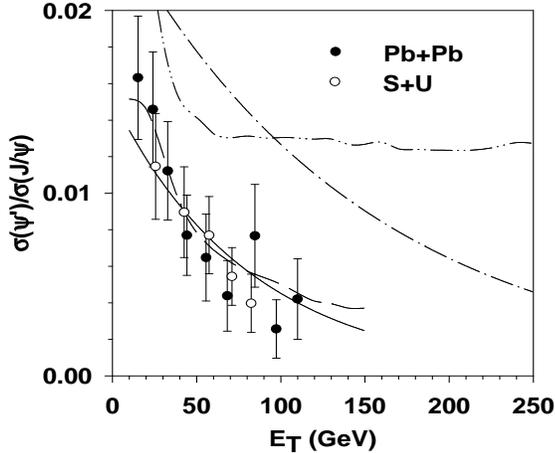,height=10cm,width=8cm}
\end{minipage}
\hspace{\fill}
\begin{minipage}[t]{80mm}
\vspace{-10.8cm}
\caption{  $E_T$  dependence  of  the  ratio,  $\psi\prime$  over
$J/\psi$, in 200 GeV/c S+U and in 158 GeV/c Pb+Pb collisions.  The
solid  and  the  dashed lines are the model calculations, for Pb+Pb
collisions  with  nuclear+comover  and  nuclear+QGP   suppression,
respectively.  The  dash-dot  and  the dash-dot-dot lines are the
predictions for the ratio at RHIC  energy,  with  nuclear+comover
and nuclear+QGP suppression respectively.}
\end{minipage}
\hspace{\fill}
\end{figure}

\vspace{-46mm}
In  Fig.1c  and  1d,  the  dashed  line is the ratio obtained with
nuclear+comover suppression, with $\sigma_{co}$=8 mb. The comover
scenario fits the $E_T$ dependence of $\psi\prime$ in S+U and  in
Pb+Pb  collisions  reasonably well. The QGP scenario, as expected,
fails to explain the S+U data but describe the  Pb+Pb  data  well,
with  the  threshold  density,  $n_c$=2.8  $fm^{-2}$. However, as
nuclear plus comover suppression also explain the data, it is not
possible to conclude positively about the formation of  QGP  from
the $E_T$ dependence of $\psi\prime$ suppression.

\section{CENTRALITY DEPENDENCE OF $\psi\prime$ OVER $J/\psi$}

Gupta  and  Satz  \cite{gupta}
proposed the ratio of $\psi\prime$  over  $J/\psi$, 
as a signal of QGP. The proposal  follows  from
the   simple   observation  that  in  a  QGP   both the $J/\psi$ and $\psi\prime$  will  be  melted.
Consequently,  the ratio $\sigma(\psi\prime)/\sigma(J/\psi)$ will
remain constant with $E_T$. Otherwise, the ratio will continually
fall  with  $E_T$,  as  $\psi\prime$  are  more  suppressed  than
$J/\psi$.                     The                    experimental
$\frac{\sigma(\psi\prime)}{\sigma(J/\psi)}$ in S+U and  in  Pb+Pb
collisions  are  shown  in  Fig.2 . For S+U collisions, the ratio
fall  continuously  with  $E_T$.  QGP  is  not  formed   in   the
collisions.  For the Pb+Pb collisions, the ratio shows a tendency
of saturation beyond 70 GeV. In Fig.2, the solid and dashed lines
are  the  ratio  for  Pb+Pb collisions in the nuclear+comover and
nuclear+QGP scenario. We note that even at large $E_T$, difference
between two model calculations are small.  Even  if  there  is  a
phase   transition,  it  will  be  difficult  to  reach  definite
conclusion at SPS energy.

At  RHIC,  the  situation  is  better.  At  RHIC  hard scattering
(proportional to number of  binary  collisions)  can  occur.  Our
prediction,  with  37\% hard scattering \cite{kh01} for the ratio
is shown in Fig.2. With  nuclear+comover absorption, the ratio 
decreases continually with $E_T$. In contrast, with nuclear+QGP suppression, the
ratio remain constant for $E_T >$ 70 GeV. The difference,  between
the two model predictions, is also large.

\section{SUMMARY AND CONCLUSIONS}

To  conclude,  we  have  analyzed  the  centrality  dependence of
$J/\psi$  and  $\psi\prime$  suppression  in  S+U  and  in  Pb+Pb
collisions.  It  was  shown that while the $J/\psi$ suppression is
well explained in the QCD based  nuclear  absorption  model,  the
model could not explain the centrality dependence of $\psi\prime$
suppression.   $\psi\prime$'s   require  additional  suppression,
either due to QGP formation or due  to  comovers,  two  scenarios
could   not  be  distinguished.  We  then  show  that  the  $E_T$
dependence of the  ratio  of  $\psi\prime$  over  $J/\psi$  could
possibly signal the deconfining phase transition at RHIC energy.

\end{document}